\documentclass[aps,prd,preprint,tightenlines,superscriptaddress]{revtex4}
\usepackage{amsmath,amssymb}
\usepackage{colordvi}
\usepackage{amsfonts}
\usepackage{enumerate}
\usepackage{slashed}
\usepackage{color}
\usepackage{xcolor}

\begin{document}

\title{Transverse Polarization of the Nucleon in Parton Picture}
\author{Xiangdong Ji}
\affiliation{INPAC, Department of Physics, and
Shanghai Key Lab for Particle Physics and Cosmology, Shanghai Jiao Tong University, Shanghai, 200240, P. R. China}
\affiliation{Center for High-Energy Physics, Peking University, Beijing, 100080, P. R. China}
\affiliation{Maryland Center for Fundamental Physics, University of Maryland, College Park, Maryland 20742, USA}
\author{Xiaonu Xiong}
\affiliation{Center for High-Energy Physics, Peking University, Beijing, 100080, P. R. China}
\affiliation{Nuclear Science Division, Lawrence Berkeley
National Laboratory, Berkeley, CA 94720, USA}
\author{Feng Yuan}
\affiliation{Nuclear Science Division, Lawrence Berkeley
National Laboratory, Berkeley, CA 94720, USA}
\date{\today}
\vspace{0.5in}
\begin{abstract}
The proton's transverse polarization structure is examined in terms of the
Lorentz-covariant Pauli-Lubanski vector in QCD. We find
that there are contributions from leading, subleading, and next-to-subleading
partonic contributions in the light-front system of coordinates. The subleading
and next-to-subleading contributions are related to the leading one through
Lorentz symmetry. And the leading contribution obeys a simple partonic
angular momentum sum rule that gives a clear physical interpretation
to a relation known previously.

\end{abstract}

\maketitle

\section{Introduction}

The proton spin structure has been one of the main focuses in
hadron physics research in the last two decades~\cite{Boer:2011fh}. The major goal is to understand
how the spin is decomposed into different components which can be interpreted
as the contributions from its constituents, quarks
and gluons. In the past, the studies are mainly focused on
the proton helicity in the set-up in which the proton travels
in the $z$-direction, with its spin polarization
along the same direction. In this case, the proton helicity
is related to the $z$-component of the angular momentum (AM) operator
$J_z$. In Ref.~\cite{Jaffe:1989jz}, Jaffe and Manohar have constructed
a gauge-dependent helicity sum rule by investigating the AM density tensor
from the basic degrees of freedom in quantum chromodynamics
(QCD). A gauge-invariant decomposition
of the proton helicity was advocated by one of the authors~\cite{Ji:1996ek},
which emphasizes the experimental accessibility of individual contributions.
The relevant generalized parton distributions
(GPDs) can be measured through deeply virtual Compton scattering
(DVCS) processes in lepton-nucleon collisions.

Beside the gauge invariance, it has also been shown that the helicity decomposition
is frame independent~\cite{Ji:1997pf}. If so, it works for the
spin-decomposition in the rest frame. Then by rotational invariance, it shall also work for the
transverse spin. However, this is not so simple. Transverse AM
does not commute with the boost operator; therefore, the partonic structure
for the transverse polarization must be different from that of the helicity.
So far, there have been confusing and conflict statements on the
transverse spin sum rule in the literature. In Ref.~\cite{Bakker:2004ib},
a transverse spin sum rule was proposed which includes the contribution from
the quark transversity distributions. However, we know that
the quark transversity is a chiral-odd object~\cite{Jaffe:1991ra}, the proposed
sum rule is in direct contradiction with the chiral-even property of the nucleon
spin and AM. Meanwhile,
an impact parameter space description for the GPDs was used by Burkardt to study the transverse
AM and spin sum rule~\cite{Burkardt:2002hr}.
A similar GPD spin sum rule was derived, where the discussions are
restricted to zero residual momentum in the infinite momentum frame (IMF) and an additional contribution
has to be included~\cite{Burkardt:2005hp}.  More recently, Leader proposed
another transverse spin sum rule~\cite{Leader:2011cr} which is different from the GPD spin sum
rule, and differs from Burkardt's derivation in IMF
as well. The current state calls for a thorough investigation
for the sum rule for the transverse polarization, and this is the
main goal of the present paper.

In order to obtain the boost-invariant spin sum rule for a moving nucleon
(along the $\hat z$ direction), we construct the polarization through the
Lorentz-covariant Pauli-Lubanski vector. Various
components of the AM tensity tensor are
found to contribute to the transverse polarization.
In light-front coordinates, the contributions correspond to
leading (twist-two), subleading (twist-three), and next-to-subleading (twist-four)
parton physics. However, due to Lorentz symmetry, all twist-three and four
contributions are related to the leading one. Thus we establish that the
sum rule derived in~\cite{Ji:1997pf} is actually a frame
independent, leading-twist, partonic sum rule for transverse polarization.
A brief summary of our results has appeared in~\cite{Ji:2012sj}.

\section{Spin Structure of the Nucleon and Frame Dependence}

In this section, we consider carefully what it means by the spin structure
of the nucleon and its frame-dependence. From this discussion, it becomes
clear that how we proceed with studying the partonic interpretation
of the transverse polarization.

It is easiest to discuss the spin structure in the nucleon's
rest frame where $\vec{P}=0$ and the nucleon is polarized in a certain
axis chosen as the $z$-direction. We have,
\begin{equation}
           J_z \left|\vec{P}=0, \frac{1}{2}\right\rangle = \frac{1}{2} \left|\vec{P}=0, \frac{1}{2}\right\rangle \ .
\end{equation}
The angular momentum projection $J_z$ in QCD can be decomposed into a sum of gauge-invariant,
local operators
\begin{equation}
   J_z= \sum_i J_z^i \ .
\end{equation}
One can take expectation values
of $J_z^i$ in the above state, and thereby obtains a sum rule or a decomposition
of $1/2=\sum_i \langle J_z^i\rangle$.  The individual contributions
$\langle J_z^i\rangle$ can be calculated
in lattice QCD or nucleon models without recoursing to partons.
Note that in general, the operator $J_z^i$ is not conserved and hence
is renormalization scale-dependent. And it does not obey the commutation relation of the full
charge $J_z$. This is the price one has to pay in quantum field theory when
discussing about the spin structure, or the structure of any other
conserved quantity, such as mass and momentum.

To study the frame-dependence of the spin structure,
it is desirable to boost the nucleon to a finite momentum. This
is easily done in the $z$-direction because the boost operator $K_z$ commutes
with $J_z$,
\begin{equation}
    [K_z, J_z]=0 \ .
\end{equation}
Thus Eq.(1) is unchanged for a finite or infinite $P_z$. On the other hand,
It is not so clear that  $\langle J_z^i\rangle$ is independent of $P_z$. It was shown
~\cite{Ji:1997pf} that so long $J_z^i$ is defined from a piece of
the AM density $M^{\mu\alpha\beta}_i$ with the same
Lorentz transformation property as the whole tensor itself,
the answer is affirmative.

To study parton physics related to $\langle J_z^i\rangle$, one has
to boost the nucleon to IMF.
Alternative to such boosts,
one can define the operators in the light-front
coordinate where the space and time undergo a special
transformation
$\xi^\pm = (\xi^0 \pm \xi^3)/\sqrt{2}$. In this case, the charge
of a symmetry is defined from the + component of a current, rather than
the 0 component. The parton picture
of $\langle J_z^i \rangle$ emerges when the operators
are interpreted in the parton degrees of freedom, in particular,
when related to some parton distributions.

In Ref. \cite{Ji:1996ek}, a sum rule has been derived for $\langle J_z^{q,g}\rangle$
from the GPD's $E^{q,g}$ and $H^{q, g}$,
\begin{equation}
   \langle J_z^{q,g}\rangle = \frac{1}{2} \int dx x [H^{q,g}(x,\eta=0,t=0)
   + E^{q,g}(x,\eta=0,t=0)] \ .
\end{equation}
This result was derived from the matrix elements of the
QCD energy momentum tensor $T^{\mu\nu}$, and has in principle no direct bearing on
the partonic interpretation of the AM contributions themselves.
Therefore, it sometimes is also called a relation instead of a sum rule
in the sense that it is not clear the integrand representing the AM density of the partons.
In particular, it has no immediate
connection with the partonic contributions to the nucleon helicity.

However, there have been some early attempts to correlate the above sum rule with a parton
picture. In \cite{Hoodbhoy:1998yb}, the parton AM density
$J(x)=(1/2)x(q(x) + E(x, 0, 0))$ was motivated from the generalization
of the AM density tensor $M^{\mu\alpha\beta}$.  The quark orbital angular
momentum (OAM) contribution to the nucleon helicity $L_q(x)$ has been identified
as $J(x) - \Delta\Sigma(x)/2$ with explicit parton OAM operators,
where $\Delta\Sigma(x)$ is the quark helicity density.
In \cite{Burkardt:2002hr}, it is found that the sum rule derived
in \cite{Ji:1996ek} has a more natural connection with the parton
contribution to the transverse spin. This observation
closely follow from the derivation of the sum rule as a spin-flip nucleon
matrix element, and points to that the transverse spin
has a more natural partonic interpretation than the longitudinal one.
Transverse spin parton sum rules were also considered
by Leader and collaborators \cite{Bakker:2004ib,Leader:2011cr}. However, a full understanding of the parton structure
of the transverse polarization deems rather complicated because
\begin{equation}
   [J_{x, y}, K_z]\ne 0   \ .
\end{equation}
The transverse AM operators do not commute with the boost operators,
therefore, a parton picture for the transverse spin AM itself will depend on
the momentum of the nucleon.

To understand properly the spin of a relativistic particle in the way consistent with
special theory of relativity, one needs to start with the covariant
spin four-vector $W_\mu$, the so-called Pauli-Lubanski vector,
\begin{equation}
W_\mu=-\frac{1}{2}\epsilon_{\mu\nu\rho\sigma}P^\nu J^{\rho\sigma}\ ,\label{pl}
\end{equation}
where $J^{\rho\sigma}$ is the angular momentum tensor, and in the light-front
coordinates, can be calculated from the angular momentum density
$M^{\mu\nu\lambda}$,
\begin{equation}
J^{\rho\sigma}=\int d\xi^-d^2\xi_\perp M^{+\rho\sigma}(x) \ .
\end{equation}
A nucleon state with momentum $P_\mu$ and polarization $S_\mu$
(which is usually normalized as $S^2=-M_N^2$), $|PS\rangle$
is an eigenstate $W^\mu \cdot S_\mu$,
\begin{equation}
    (W^\mu \cdot S_\mu)\left|PS\right\rangle = \frac{1}{2} (-M_N^2) \left|PS\right\rangle \ .
\end{equation}
Therefore, to study the transverse polarization, one has to start with
the Pauli-Lubanski vector along the transverse direction,
\begin{equation}
W^\perp=\epsilon^{-+\perp\sigma}\left(P^+J^{-\sigma}-P^-J^{+\sigma}\right) \ , \label{wt}
\end{equation}
which not only involves the AM $J^{+\sigma}$, but also the
boost operator $J^{-\sigma}$. Therefore, a frame-independent
transverse polarization is achieved through the presence of the boost generator.
And only in the rest frame of the nucleon, the boost operator drops out, $W^\perp$
reduces to the transverse AM. Therefore, a study of transverse polarization
cannot avoid discussing the matrix element of the boost operator.

\section{Structure of Transverse Polarization from Energy-Momentum Tensor}

From the general discussions in the previous section, we know that to obtain
a transverse polarization sum rule, we have to start from the Pauli-Lubanski vector.
In this section, we will study the structure of transverse polarization
from the matrix elements of the AM density and energy-momentum tensor.
We assume the AM density $M^{\mu\nu\lambda}$ can be decomposed into a sum of
different parts,
\begin{equation}
   M^{\mu\nu\lambda} = \sum_i M_i^{\mu\nu\lambda} \ .
\end{equation}
The following discussion is applicable to individual part, $M_i^{\mu\nu\lambda}$,
\begin{equation}
M^{\mu\nu\lambda}_i(x)=x^\nu T^{\mu\lambda}_i-x^\lambda T^{\mu\nu}_i \ ,
\end{equation}
which can be further defined
from the quark and gluon contributions of the energy-momentum tensor in QCD,
\begin{eqnarray}
T^{\mu\nu}=T^{\mu\nu}_q+T^{\mu\nu}_g \ ,\label{tmu}
\end{eqnarray}
where
\begin{eqnarray}
T^{\mu\nu}_q&=&\frac{1}{2}\left[\bar\psi\gamma^{(\mu}i\overrightarrow{D}^{\nu)}\psi+\bar\psi\gamma^{(\mu}i\overleftarrow{D}^{\nu)}\psi\right]\nonumber\\
T^{\mu\nu}_g&=&\frac{1}{4}F^2g^{\mu\nu}-F^{\mu\alpha}{F^{\nu}}_{\alpha} \ .\label{tmuqg}
\end{eqnarray}
Thus we ultimately relate the expectation value of the Pauli-Lubanski
operator in the transversely-polarized nucleon state
to the matrix elements of the energy-momentum tensor.

Following Ref.~\cite{Jaffe:1989jz}, we define the off-forward matrix element of $M^{\mu\nu\lambda}$
in the nucleon state
\begin{equation}
{\cal M}^{\mu\nu\lambda}_i(k)=\int d^4x e^{ik\cdot x} \langle P'S'|M^{\mu\nu\lambda}_i(x)|PS\rangle \ ,
\end{equation}
In the end, we will take $P'=P$ and $S'=S$. After subtracting the total derivative, we
obtain
\begin{eqnarray}
{\cal M}^{\mu\nu\lambda}_i(k)=-i(2\pi)^4\delta^{4}(k+P'-P)\left\{
\frac{\partial}{\partial k_\nu}\left[ \langle P'S|T^{\mu\lambda}_i(0)|PS\rangle\right]-
\left(\nu\leftrightarrow \lambda\right) \right\}\ .
\end{eqnarray}
General expressions for ${\cal M}^{\mu\nu\lambda}$ has been discussed in Ref.~\cite{Jaffe:1989jz}.

To complete the calculation, we need the parameterization for the matrix element of
the energy-momentum tensor~\cite{Ji:1996ek},
\begin{eqnarray}
\langle P'S|T^{\mu\nu}_i(0)|PS\rangle&=&\bar U(P') \left[A_i(\Delta^2)\gamma^{(\mu}\bar P^{\nu)}
+B_i(\Delta^2)\frac{\bar P^{(\mu}i\sigma^{\nu)\alpha}\Delta_\alpha}{2M_N}\right.\nonumber\\
&&\left.+C_i(\Delta^2)\frac{\Delta^\mu\Delta^\nu-g^{\mu\nu}\Delta^2}{M_N}
+\bar C_i(\Delta^2)M_Ng^{\mu\nu}\right]U(P ) \  , \label{energy}
\end{eqnarray}
where $\bar P=(P+P')/2$, $\Delta=P'-P$, and $A$, $B$, $C$ and $\bar C$
are form factors. To calculate the first-order derivative, we expand the above
result to the linear term of $\Delta^\alpha$~\cite{Jaffe:1989jz}. We can immediately
drop out the contribution from the $C$ form factor because it is proportional
to $\Delta^2$. As we shall see, the contributions from $\bar C_i$
form factors is related to the twist-four contribution which
cancels between quarks and gluons.

Therefore, in the following discussions, we only keep $A$ and $B$
form factors in the above equation.
To further simplify this equation, we apply the Gordon identity,
to get
\begin{eqnarray}
\langle P'S|T^{\mu\nu}_i(0)|PS\rangle=\bar U(P') \left[A_i(\Delta^2)\frac{\bar P^\mu\bar P^{\nu}}{M_N}
+\left(A_i(\Delta^2)+B_i(\Delta^2)\right)\frac{\bar P^{(\mu}i\sigma^{\nu)\alpha}\Delta_\alpha}{2M_N}\right]U(P ) \  .\label{energytensor}
\end{eqnarray}
It is clear that the second term of Eq.~(\ref{energytensor}) has $\Delta^\alpha$ dependence
explicitly, whose contribution can be easily evaluated.
On the other hand, the first term of Eq.~(\ref{energytensor}) is a little involved.
In most cases, it does not contain a linear term of $\Delta_\alpha$. This is the case
if the nucleon is in the rest frame. It is also true for the longitudinal polarization
in the moving frame (along $\hat z$ direction) with $P_z\neq 0$. However, it does contribute
to a linear term for the transverse polarization with nonzero $P_z$, and its contribution
depends on $P_z$. This indicates that the contribution of this term is not boost invariant.
This was first realized in Ref.~\cite{Bakker:2004ib}.

A transverse polarization sum rule starts from
the expression of the transverse component of the
Pauli-Lubanski vector. In anticipation to
studying the partonic structure, we use the light-front coordinates. Therefore, we have,
\begin{equation}
W^{\perp}_i=\epsilon^{-+\perp\sigma}\left(P^+J^{-\sigma}_i-P^-J^{+\sigma}_i\right) \ , \label{wt}
\end{equation}
which involves ${\cal M}^{++\perp}_i$ and ${\cal M}^{+-\perp}_i$,
\begin{eqnarray}
M^{++\perp}_i(x)&=&x^+T^{+\perp}_i-x^\perp T^{++}_i \ ,\\
M^{+-\perp}_i(x)&=&x^-T^{+\perp}_i-x^\perp T^{+-}_i\ .
\end{eqnarray}
From the above equation, we need the matrix elements the energy-momentum
tensors: $T^{++}_i$, $T^{+\perp}_i$, and $T^{+-}_i$. Clearly, $T^{++}_i$ has parton density
interpretation as discussed in Ref.~\cite{Ji:2012sj}, whereas the rest two
will involve twist-three and four effects which do not. This indicates that a complete
picture of the transverse polarization in partons is complicated.
This situation is quite normal in light-front coordinates. For example,
consider any vector $V^\mu$, it is the plus component $V^+$
has the simple partonic interpretation, whereas $V^\perp$ and $V^-$ are subleading
and the next-to-subleading in light-cone power counting. When discussing the magnitude
of the vector, one may focus on the leading component only and using Lorentz symmetry
to simplify the parton picture for the other components when possible.

Indeed as we shall see, in the case of transverse Pauli-Lubanski vector,
one can use the symmetry argument to relate the subleading contributions
to the leading one to develop a simple parton picture.
To study the contributions from the different components,
we first consider
\begin{eqnarray}
\langle P'S'|T^{++}_i(0)|PS\rangle&=&\bar U(P') \left[A_i(\Delta^2)\frac{\bar P^+\bar P^{+}}{M_N}
+\left(A_i(\Delta^2)+B_i(\Delta^2)\right)\frac{\bar P^{+}i\sigma^{+\alpha}\Delta_\alpha}{2M_N}\right]U(P )\ ,\label{pp}\\
\langle P'S'|T^{+\perp}_i(0)|PS\rangle&=&\bar U(P') \left[
\left(A_i(\Delta^2)+B_i(\Delta^2)\right)\frac{\bar P^{(+}i\sigma^{\perp)\alpha}\Delta_\alpha}{2M_N}\right]U(P )\ , \\
\langle P'S'|T^{+-}_i(0)|PS\rangle&=&\bar U(P') \left[A_i(\Delta^2)\frac{\bar P^+\bar P^{-}}{M_N}
+\left(A_i(\Delta^2)+B_i(\Delta^2)\right)\frac{\bar P^{(+}i\sigma^{-)\alpha}\Delta_\alpha}{2M_N}\right]U(P )
 \  .\label{pm}
\end{eqnarray}
It is easy to see that the first term of Eq.~(\ref{pp}) cancels out the first term of Eq.~(\ref{pm}) in
the contribution to $W_\perp$ in Eq.~(\ref{wt}). This cancelation
allows one to see a simple boosting property of the $W_\perp$ at different
longitudinal momentum. In the following discussion, we will neglect both terms.
We further notice that the second term of Eq.~(\ref{pm}) vanishes
as well, because it is proportional to $P^+P^--P^-P^+$ due the antisymmetric
nature of $\sigma^{\mu\nu}$. A possible contribution from $\bar C$ form factor
cancel between quarks and gluons because $\bar C_q(0) +\bar C_g(0) = 0$.
Therefore, we take no contribution to $W_{\perp i}$
from the energy-momentum tensor $T^{+-}_i$.

We are left with contributions from $T^{++}_i$ and $T^{+\perp}_i$. Expanding
these results to the linear term of $\Delta_\alpha$, we obtain,
\begin{eqnarray}
\langle P'S'|T^{++}_i(0)|PS\rangle&=&\left[\frac{A_i(0)+B_i(0)}{2}\right]\frac{2(\bar P^{+})^2}{M_N^2}\epsilon^{-+\alpha\beta}(i\Delta_\alpha) S_\beta\  ,\\
\langle P'S'|T^{+\perp}_i(0)|PS\rangle&=&\left[\frac{A_i(0)+B_i(0)}{2}\right]\frac{(\bar P^{+})^2}{M_N^2}\epsilon^{-\perp\alpha\beta}(i\Delta_\alpha) S_\beta\  ,
\end{eqnarray}
where we have dropped the first term of Eq.~(\ref{pp}) as we commented above.

Applying the above expansion results, we obtain the angular momentum
tensor,
\begin{eqnarray}
{\cal M}^{++\perp}_i&=&(2\pi)^4\delta^{4}(0)\left[\frac{A_i(0)+B_i(0)}{2}\frac{3}{2}\right]\frac{2(\bar P^{+})^2S^{\perp'}}{M_N^2} \ , \\
{\cal M}^{+-\perp}_i&=&-(2\pi)^4\delta^{4}(0)\left[\frac{A_i(0)+B_i(0)}{2}\frac{1}{2}\right]\frac{2\bar P^{+}\bar P^-S^{\perp'}}{M_N^2} \ ,
\end{eqnarray}
where $S^{\perp'}=\epsilon^{-+\perp\alpha}S^\alpha$. Substituting these
results into Eq.~(\ref{wt}), we find that
\begin{equation}
W^{\perp}_i=\frac{A_i(0)+B_i(0)}{2}S^\perp \ ,
\end{equation}
Thus the contribution $T^{+\perp}_i$ is crucial to
obtain the complete result for the transverse polarization.
To demonstrate this more clearly, we show separately the above contributions
from $T^{++}_i$ and $T^{+\perp}_i$,
\begin{eqnarray}
W^{\perp}_i |_{T^{++}_i}&=&\frac{A_i(0)+B_i(0)}{4}S^\perp \ , \\
W^{\perp}_i |_{T^{+\perp}_i}&=&\frac{A_i(0)+B_i(0)}{4}S^\perp \ .
\end{eqnarray}
Thus $T^{++}_i$ and $T^{+\perp}_i$
contribute separately 1/2 of the nucleon spin. This a simple result
of Lorentz symmetry and cannot be obtained without including
the boost operator in the Pauli-Lubanski vector.  The above result
will be used to seek partonic interpretation
of the transverse spin sum rule in the next section.

\section{Partonic Sum Rule for Transverse Polarization}

The discussion in the previous section
allows one to derive simple partonic sum rule for the transverse
polarization. Although $W_\perp$ receives contributions from
leading $T^{++}$ and subleading $T^{+\perp}$ and $T^{+-}$ contributions,
the unwanted Lorentz structure from $T^{+-}$ cancels that of
$T^{++}$ and drops out of the discussion. The contribution from
$T^{+\perp}$ is then the same as the contribution of $T^{++}$ by
Lorentz symmetry. We notice that the partonic content of $T^{+\perp}$ 
is complicated, since it involves a twist-three parton correlation which 
we will not discuss in this paper.
However, as we demonstrated in the last section, both terms contribute 
to a fixed equal amount of the transverse polarization, and thus they are 
dynamically independent. Therefore, in this section, we will focus on the 
leading twist-two part of $T^{++}$ contribution, which yields a 
frame-independent simple parton picture of the transverse polarization.

We first specialize the discussion of the previous section to the quark and gluon parts
of the energy momentum tensor,
$T_q^{++}$ and $T_g^{++}$, which contribute to the AM density,
\begin{eqnarray}
{\cal M}_{q,g}^{++\perp}|_{T^{++}}&=&(2\pi)^4\delta^{(4)}(0)\left[\frac{A_{q,g}(0)+B_{q,g}(0)}{2}\right]\frac{2(\bar P^{+})^2S^{\perp'}}{M_N^2}  \ ,
\end{eqnarray}
where the form factors have been defined before. Their contributions
to the transverse polarization sum is,
\begin{equation}
W^\perp_{q,g}|_{T^{++}}=\frac{A_{q,g}(0)+B_{q,g}(0)}{4}S^\perp \ .  \label{tranpp}
\end{equation}
The explicit expression for quark and gluon energy-momentum tensors $T^{++}_{q,g}$ is,
\begin{eqnarray}
T^{++}_q&=&\frac{1}{2}\left[\bar \psi\gamma^+\overrightarrow{D}^+\psi+\bar\psi\overleftarrow{D}^+\gamma^+\psi\right]\ ,\\
T^{++}_g&=&F^{+i}F^{+i} \ ,
\end{eqnarray}
where $i=1,2$.

To explore the partonic picture for these contributions, we first
calculate the parton momentum density when the nucleon is
in transverse polarization,
\begin{equation}
   \rho_q^{+} (x, \xi,S^\perp) = x\int \frac{d\lambda}{4\pi} e^{i\lambda x}
     \left\langle PS^\perp\left|\overline \psi\left(-\frac{\lambda n}{2},\xi_\perp\right)\gamma^+\psi\left(\frac{\lambda n}{2},\xi_\perp\right)
     \right|PS^\perp\right\rangle \ ,
\end{equation}
where $\xi_\perp$ denotes the spatial transverse coordinates, and
$n$ is an adjoint vector with $n^2=0$ and $n\cdot P=1$. This function
describes the longitudinal momentum distribution of quarks carrying the momentum fraction
$x$ and at the coordinate $\xi_\perp$ in the transverse plane. Nominally
because of the translational symmetry in the transverse plane, $
\rho_q^+$ shall have no dependence on $\xi_\perp$. However,
there is a subtle distribution term in the mathematical sense which
can only be revealed from the forward limit of an off-forward matrix element. 
Using the GPD's defined in
\begin{eqnarray}
&&\int \frac{d\lambda}{2\pi}e^{i\lambda x}\left\langle P'|\bar \psi\left(-\frac{\lambda n}{2}\right)\gamma^+\psi\left(\frac{\lambda n}{2}\right)
|P\right\rangle \nonumber\\
&&~~
= H_q(x,\eta,\Delta)\bar U(P')\gamma^+U(P )+E_q(x,\eta,\Delta)\bar U(P')\frac{i\sigma^{+\alpha}\Delta_\alpha}{2M_N}U(P ) \ ,
\end{eqnarray}
where $\eta$ is the skewness parameter, 
a transverse-polarization dependent term appears
\begin{equation}
   \rho_q^{+} (x, \xi,S^\perp)/P^+= xH_q(x,0,0)+\left[\frac{xH_q(x,0,0)+xE_q(x,0,0)}{2}\right]\lim_{\Delta_\perp\rightarrow 0}\frac{S^{\perp'}}{M^2}
   \partial^{\perp_\xi} e^{i\xi_\perp\Delta_\perp} \ ,
   \end{equation}
where we have neglected a contribution from the scalar term in the Gordon identity.
This term shall be canceled by a similar contribution from the twist-four operators
as we discussed before.

In a sense, $\rho^+$ provide the joint distribution of partons with  longitudinal momentum $x$ 
and transverse coordinate $\xi_\perp$. When integrating out $\xi_\perp$, the second term drops 
out, and we recover the usual momentum distribution of the quarks. On the other hand, if we 
integrate over $\xi$ with a weight $\xi^\perp$, the first term drops out, and the second term 
leads to the AM contribution from quarks with fixed longitudinal momentum $x$,
\begin{eqnarray}
    {\cal M}^{++\perp}_q(x)|_{T^{++}} &=&P^+(2\pi)^2\delta^{(2)}(0)
   \int d^2\xi \xi^{\perp} \rho_q^{+}(x, \xi,S^\perp)\nonumber\\
   & =& (2\pi)^4\delta^{(4)}(0)\frac{x}{2}\frac{H_q(x,0,0)+E_q(x,0,0)}{2} \frac{2(\bar P^{+})^2S^{\perp'}}{M_N^2} \ ,
\end{eqnarray}
and the corresponding contribution to the transverse polarization is
\begin{eqnarray}
W^\perp_{q}(x)|_{T^{++}}&=&\frac{M_N^2}{2P^+(2\pi)^2\delta^{(2)}(0)}\int d^2\xi \xi^{\perp'} \rho_q^{+}(x, \xi,S^\perp)\nonumber\\
&=&S^\perp\frac{x}{2}\frac{H_q(x,0,0)+E_q(x,0,0)}{2}\ .
\label{wtq}
\end{eqnarray}
Therefore, we have shown that quarks with longitudinal momentum $x$ 
contribute $S_q^\perp(x) = (x/2)(E_q(x,0,0)+H_q(x,0,0)$ to the transverse 
polarization. By integrating over $x$, we obtain the total contribution 
\begin{equation}
  S_q^\perp = \frac{1}{2}\int dx x (H_q(x,0,0) + E_q(x,0,0)) \ . 
\end{equation}
as a partonic sum rule for the
transverse polarization.

Similarly, for the gluons, we define the longitudinal momentum density,
\begin{equation}
   \rho_g^{+} (x, \xi_\perp,S^\perp) = \int \frac{d\lambda}{2\pi} e^{i\lambda x}
     \langle PS^\perp|F^{+i}(-\frac{\lambda n}{2},\xi_\perp)F^{+i}(\frac{\lambda n}{2},\xi_\perp)|PS^\perp\rangle \ ,
\end{equation}
then the gluon parton of momentum $x$ contributes to the transverse polarization
density, 
\begin{eqnarray}
W^\perp_{g}(x)|_{T^{++}}&=&\frac{M_N^2}{2P^+(2\pi)^2\delta^{(2)}(0)}\int d^2\xi \xi^{\perp'} \rho_g^{+}(x, \xi,S^\perp)\nonumber\\
&=&S^\perp\frac{x}{2}\left(H_g(x,0,0)+E_g(x,0,0)\right)\ ,
\label{wtg}
\end{eqnarray}
The total contribution is a partonic sum rule, 
\begin{equation}
  S_g^\perp = \frac{1}{2}\int dx x (H_g(x) + E_g(x)) \ , 
\end{equation}
which is just the GPD sum rule in Eq. (4). 

Eqs.~(\ref{wtq},\ref{wtg}) are the main results of our paper. They are derived from the
angular momentum density in QCD with the Lorentz symmetry argument for the
transverse polarization decomposition for nucleon state. They provide an intuitive 
picture for the nucleon transverse polarization from the quarks and gluons.

Comparing with Leader's transverse spin sum rule derived recently~\cite{Leader:2011cr}, 
we found that his result is frame dependent. As we discussed before, this frame dependence arises
from the non-commutativity of the transverse AM and longitudinal boost. Therefore, 
a frame-independent parton picture does not exist for the transverse AM but rather
the Pauli-Lubanski spin vector. Comparing with Burkardt's result~\cite{Burkardt:2005hp}, 
our derivation here dispenses with the wave-packet construction and is valid for any 
residual momentum of the nucleon in IMF, not just the rest frame.

\section{Summary}

In conclusion, we have examined the transverse spin structure for the nucleon
by a detailed study of the angular momentum density tensor. By constructing
transverse polarization through the Lorentz-covariant Pauli-Lubanski vector, we derived
a sum rule that satisfies the boost invariance, and is consistent with the
GPD spin sum rule derived early. We find that the leading contribution to the
the transverse AM has a simple partonic AM density interpretation.

We thank M.~Burkardt and E.~Leader for the comments. 
This work was partially supported by the U.
S. Department of Energy via grants DE-FG02-93ER-40762
and DE-AC02-05CH11231 and
a grant (No. 11DZ2260700) from the Office of Science and Technology in Shanghai Municipal Government.


\begin{thebibliography}{99}

\bibitem{Boer:2011fh}
  D.~Boer, 
   {\it et al.},
  arXiv:1108.1713 [nucl-th].

\bibitem{Jaffe:1989jz}
  R.~L.~Jaffe and A.~Manohar,
  Nucl.\ Phys.\ B {\bf 337}, 509 (1990).


\bibitem{Ji:1996ek}
  X.~Ji,
  Phys.\ Rev.\ Lett.\  {\bf 78}, 610 (1997).



\bibitem{Ji:1997pf}
  X.~Ji,
  Phys.\ Rev.\ D {\bf 58}, 056003 (1998).

\bibitem{Bakker:2004ib}
  B.~L.~G.~Bakker, E.~Leader and T.~L.~Trueman,
  Phys.\ Rev.\ D {\bf 70}, 114001 (2004).

\bibitem{Jaffe:1991ra}
  R.~L.~Jaffe and X.~-D.~Ji,
  Nucl.\ Phys.\ B {\bf 375}, 527 (1992).

\bibitem{Burkardt:2002hr}
  M.~Burkardt,
  Int.\ J.\ Mod.\ Phys.\ A {\bf 18}, 173 (2003).
\bibitem{Burkardt:2005hp}
  M.~Burkardt,
  Phys.\ Rev.\ D {\bf 72}, 094020 (2005).

\bibitem{Leader:2011cr}
  E.~Leader,
  arXiv:1109.1230 [hep-ph].

\bibitem{Ji:2012sj}
  X.~Ji, X.~Xiong and F.~Yuan,
  arXiv:1202.2843 [hep-ph].

\bibitem{Hoodbhoy:1998yb}
  P.~Hoodbhoy, X.~-D.~Ji and W.~Lu,
  Phys.\ Rev.\ D {\bf 59}, 014013 (1999)
  [hep-ph/9804337].



\end{thebibliography}
\end{document}